\documentclass[%
reprint, 
 superscriptaddress,
showpacs,preprintnumbers,
 amsmath,amssymb,
 aps,
]{revtex4-1}

\usepackage{graphicx}
\usepackage{dcolumn}
\usepackage{bm}
\usepackage{color}

\usepackage{cancel}
\usepackage{multirow}
\usepackage{amsmath}
\usepackage{enumitem}

\def\met{\cancel{E}_T}
\newcommand{\Hpm}{H_5^{\pm}}
\newcommand{\Hpmpm}{H_5^{\pm\pm}}
\newcommand{\eFwd}{e_{\rm fwd}}
\newcommand{\lZ}[1]{l_{\rm{z,#1}}}
\newcommand{\jFwd}{j_{\rm fwd}}
\newcommand{\jW}[1]{j_{\rm{w,#1}}}

\newcommand{\MHpm}{M_{H_5^{\pm}}}
\newcommand{\MH}{M_{H_5}}
\newcommand{\sTheta}{\sin\theta_H}

\newcommand{\iab}{{\rm ab}^{-1}}
\newcommand{\ifb}{{\rm fb}^{-1}}

\begin{document}

\preprint{DESY 17-150}

\title{
Search for Singly Charged Higgs in Vector Boson Scattering at the ep Colliders
}

\author{Georges Azuelos}
\email{georges.azuelos@cern.ch}
\affiliation{ Universit\'e de Montr\'eal, Montr\'eal,  Canada }
\affiliation{ TRIUMF, Vancouver, Canada}

\author{Hao Sun}
\email{haosun@dlut.edu.cn}
\affiliation{ Institute of Theoretical Physics, School of Physics, Dalian University of Technology, No.2 Linggong Road, Dalian, Liaoning, 116024, P.R.China }%

\author{Kechen Wang}
\email{kechen.wang@desy.de (Corresponding author)}
\affiliation{ DESY, Notkestrae 85, D-22607 Hamburg, Germany }%
\affiliation{ Center for Future High Energy Physics, Institute of High Energy Physics, Chinese Academy of Sciences, Beijing, 100049, China }%

\date{\today}

\begin{abstract}
We search for a five-plet singly charged Higgs $\Hpm$ in the Georgi-Machacek model at the ep colliders.
The charged Higgs bosons are produced via the $ZW^{\pm}$ fusion process $p\, e^- \to j\, e^-\, \Hpm$, and decay as $\Hpm \to Z\, W^{\pm} \to (l^+ l^-)\, (jj)$. With a detector-level simulation at the FCC-eh and LHeC, a multi-variate analysis is performed to yield limits on the production cross section times branching ratio $\sigma (p\, e^- \to j e^- \Hpm) \times {\rm BR}(\Hpm \to Z\, W^\pm)$ and on the model parameter $\sTheta$ for charged Higgs masses between 200 and 1000 GeV. The effects of electron beam polarization are also investigated.
\end{abstract}

\maketitle


\section{Introduction}
\label{sec:intro}

The discovery of the Higgs boson at the Large Hadron
Collider (LHC)\cite{SMHiggs_ATLAS, SMHiggs_CMS}
is a major step towards understanding of
the electroweak symmetry breaking (EWSB)
mechanism and marks a new era in particle physics, but this is not the end of the story.
In fact, from a theoretical point of view, there is no fundamental reason
for a minimal Higgs sector, as occurs in the Standard Model (SM).
It is therefore important to consider extended scalar scenarios with higher isospin multiplets
that may also contribute to EWSB.
They could also provide a good way to generate
a Majorana mass for neutrinos through the type-II seesaw\cite{seesaw_TPYEII}
mechanism.

One such scenario is the Georgi-Machacek (GM)~\cite{Chaow} model which contains
a complex $\rm SU(2)_L$ doublet field, a real triplet field
and a complex $\rm SU(2)_L$ triplet field.
Compared with the other Higgs extended models, such as a Left-Right symmetric model\cite{ExtendedHiggs_LR}
or a Little Higgs model\cite{ExtendedHiggs_LH}, the GM model has some desirable features.
It preserves the custodial $\rm SU (2)_C$ symmetry at tree level,
keeping the electroweak $\rm \rho$ parameter close to unity.
It is thus less constrained experimentally~\cite{Godfrey:2010qb}.
After symmetry breaking, the physical fields can be organized by their transformation properties
under the custodial $\rm SU(2)_C$ symmetry into a fiveplet, a triplet, and two singlets.
One of the singlets is the SM-like Higgs, whose tree level couplings to fermions
and vector bosons may be enhanced in comparison to the SM case~\cite{Chiang:2015kka}.
The other singlet and the fiveplets couple to the electroweak gauge bosons but not to the SM quarks
at tree level, whereas the triplet couples to the quarks but not to the gauge bosons,
and thus they can be studied through different channels.
The exotic scalars belonging to the custodial fiveplets include
neutral, singly- and doubly-charged members: $\rm H^0_5, \rm H^\pm_5, \rm H^{\pm\pm}_5$.
The appearance of the custodial fiveplet particles results in a rich phenomenology.
For example, the doubly-charged Higgs boson can couple to a pair of same-sign W bosons,
through the vertex ${\rm G_{H^{\pm\pm}_5W^\mp W^\mp}} = 2 e^2 v_{\chi}/s_W^2$, proportional to a larger vacuum expectation value ($v_\chi$) than custodial symmetry allows.
This may provide additional contributions to the SM quartic gauge couplings~\cite{GM_Boundary_b}.
The neutral custodial fiveplet Higgs
couples to both WW and ZZ and its observation can be used to test the mass degeneracy of
charged and neutral scalar bosons in the GM model\cite{ILC_sinh}.
The singly charged member of the custodial fiveplet Higgs also couples to the electroweak gauge bosons
and provides a good testing ground for the detection of the $\rm H^\pm W^\mp Z$ vertex ${\rm G_{H^\pm W^\mp Z}} = - \sqrt{2} e^2 v_{\chi} / c_W s_W^2 $~\cite{Kanemura_HWZ}.
The vacuum expectation value can be parameterized as ${\rm \sin\theta_{H}}=2\sqrt{2} v_\chi/v$,
where $v$ is the SM VEV and the $\rm H^\pm W^\mp Z$ vertex is then directly proportional to
the parameter $\sin\theta_{H}$.
This vertex appears at tree level only when $\rm H^\pm$ comes
from an exotic representation such as a triplet. It is absent at tree level if $\rm H^\pm$ comes from a doublet.
Therefore, this vertex can be used to distinguish between models with singly charged Higgs bosons.

Indirect limits on the GM model parameters can be set from B-physics and precision electroweak measurements~\cite{Hartling2015}.
Based on the search for a heavy charged Higgs boson produced through vector boson fusion and decaying into $W Z$ bosons,
direct limits on the parameter $\rm \sin\theta_H$ have been obtained as a function of the charged Higgs mass by the ATLAS collaboration~\cite{Aad:2015nfa} from $pp$ collision at a centre-of-mass energy of 8 TeV.
Recently, more stringent limits have been reported by the CMS collaboration from data at $\sqrt{s}= 13$ TeV with 15.2 $\ifb$ luminosity ~\cite{GM_Boundary_H5P_CMS}.
In addition to the direct singly-charged Higgs searches, searches from doubly-charged Higgs production
may also set limits on this parameter since different charged Higgs boson states belonging to the same
multiplet are expected to be degenerate in GM model.
The CMS collaboration has reported limits from the doubly charged Higgs searches
through the electroweak production of same-sign W
boson pairs in the final state of two jets, two same-sign leptons and missing energy
with 35.9 $\rm fb^{-1}$ of integrated luminosity at a centre -of-mass energy of 13 TeV~\cite{GM_Boundary_H5PP_b}.
The sensitivity for the doubly-charged Higgs searches at $ep$ colliders has been evaluated in Ref.~\cite{GM_H5pp_haosun}.

In this paper, we evaluate the sensitivity of $ep$ colliders to measure the vertex $\rm H_5^\pm W^\mp Z$.
Based on the framework of GM model,
we perform a detector-level simulation at both LHeC and FCC-eh colliders for the signal process
$p\, e^- \to j\, e^-\, \Hpm$, produced via $WZ$ fusion and followed,
with 100\% branching ratio,
by  $\Hpm \to Z\, W^{\pm} \to (l^+ l^-)\, (jj)$
(see Fig. 1).
Although this semileptonic channel of $WZ$ decay has a relatively low branching ratio compared with the decays of $Z\, W^{\pm} \to (jj)\, (l \nu)$ or $(jj)\, (jj)$, its SM background is also relatively low and a good reconstruction of the signal is possible.
For the $e^-$ and $p$ beam energies at the LHeC and FCC-eh colliders, we consider 60 GeV$\times$ 7 TeV, and 60 GeV $\times$ 50 TeV~\cite{AbelleiraFernandez:2012cc,ehParameters}.
Although the  center-of-mass energies of  1.3 TeV and 3.5 TeV, respectively, are lower than at $pp$ colliders, the SM QCD backgrounds are much smaller and pileup jets are essentially negligible at $ep$ colliders.

The article is organized as follows.
In Sec.~\ref{sec:strategy}, we discuss the distinguishing features of the signal and present the analysis method.
In Sec.~\ref{sec:reuslts}, we give the numerical results and set limits on the GM model parameters.
We summarize and conclude in the last Sec.~\ref{sec:Summary}.


\section{Search Strategy}
\label{sec:strategy}

The chain of data simulation starts with the event generator MadGraph5\_aMC@NLO~\cite{Alwall:2014hca}. Parton showering and hadronization is then performed by Pythia~\cite{Sjostrand:2007gs}. Delphes~\cite{deFavereau:2013fsa} is used for detector simulation. The detector is assumed to have a cylindrical geometry comprising a central tracker followed by an electromagnetic and a hadronic calorimeter. The forward and backward regions are also covered by a tracker, an electromagnetic and a hadronic calorimeter. The angular acceptance for charged tracks in the pseudorapidity range of $-4.3 < \eta < 4.9$ and the detector performance in terms of momentum and energy resolution of electrons, muons and jets, are based on the LHeC detector design~\cite{AbelleiraFernandez:2012cc,LHeCDetector}.
For our simulation, 
a modified Pythia version tuned for the ep colliders and the Delphes card files for the LHeC and FCC-eh detector configurations~\cite{Delphes_cards} are used.

For the signal, we consider that only the 5-plet of the GM model is sufficiently light to be within reach of the collider while other new scalars are heavy and decoupled. The simplified model has only 2 free parameters: the mass of the 5-plet $\MHpm$ and the $\sTheta$ value.
The UFO model file from Ref.~\cite{GMModel} is utilized to simulate the signal data.

\begin{figure}[h]
\includegraphics[scale=0.14]{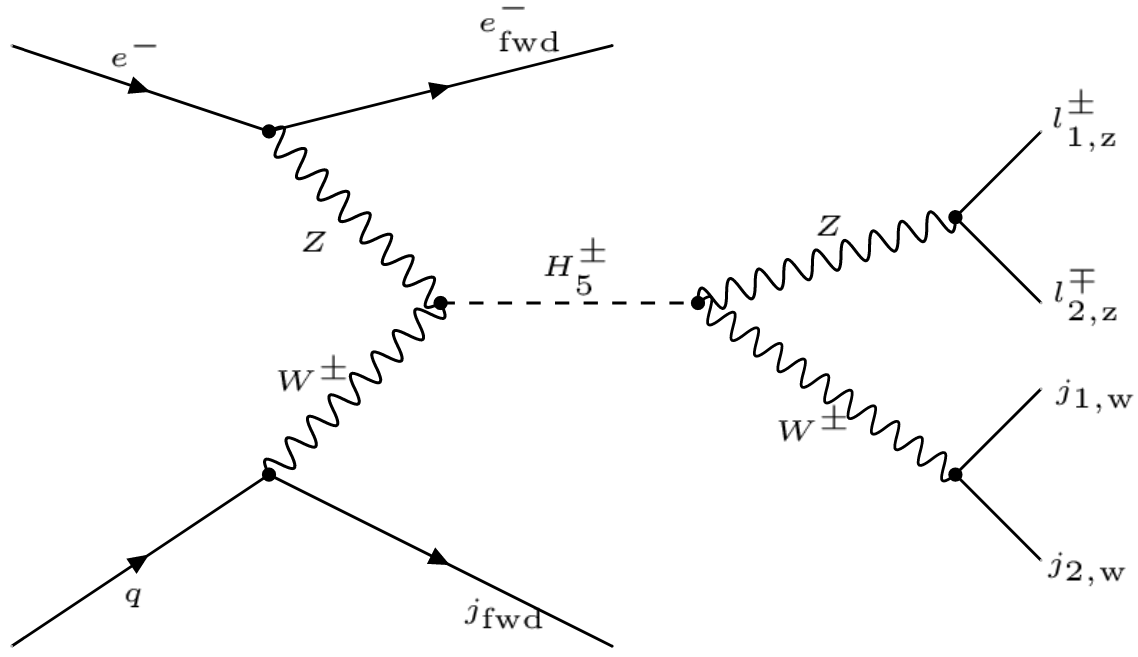}
\caption{ The Feynman diagram for the signal production of  $p\, e^- \to j\, e^-\, \Hpm$, followed by the decays of $\Hpm \to Z W^{\pm} \to (l^+l^-)\, (jj)$. }
\label{fig:diagram}
\end{figure}

Among the three leptons in the final state,
one electron $\eFwd$ is from the incoming electron beam.
The Z boson can be reconstructed from two Opposite Sign Same Flavor (OSSF) leptons $l^+l^-$ (here $l$ denotes $e$ and $\mu$ only) which are ordered in decreasing transverse momentum and are here  labeled as $\lZ1$ and $\lZ2$.
Among the three jets,
one forward jet $\jFwd$ is from the incoming proton beam.
The leading and subleading jets from the W boson decay are labeled as $\jW1$ and $\jW2$.

In Fig.~\ref{fig:crs}, we show the cross sections in fb at the FCC-eh and LHeC, for the process $p\, e^- \to j\, e^-\, \Hpm$ followed by the decays of $\Hpm \to Z W^{\pm}$, as a function of the mass $\MHpm$ and with the model parameter $\sTheta=0.5$.
Another factor of 4.1\% applies when considering the subsequent decays of $Z \to l^+l^-$ and $W^{\pm} \to jj$.
It can be seen that at the FCC-eh, the -80\% (+80\%) beam polarization increases (decreases) the production cross section by a factor of about 10\%, compared with the unpolarized beam case.

\begin{figure}[h]
\includegraphics[scale=0.13]{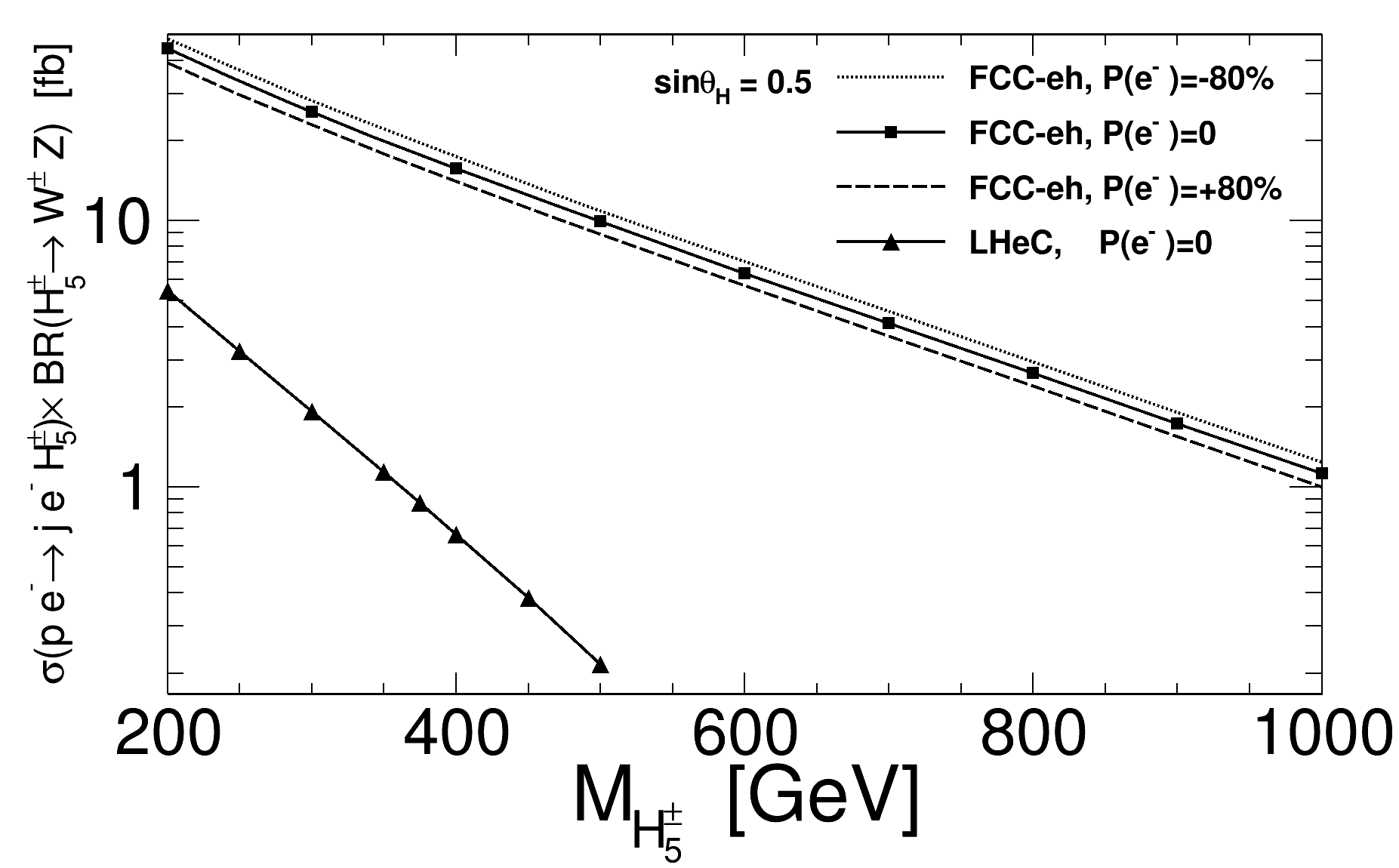}
\caption{ The production cross section times branching ratio $\sigma (p\, e^- \to j e^- \Hpm) \times {\rm BR}(\Hpm \to Z\, W^\pm) $ in fb at the FCC-eh with -80\% polarized (dotted line), unpolarized (solid line with square markers), or +80\% polarized (dashed line) electron beam, and at the LHeC with unpolarized electron beam (solid line with triangle markers) as a function of the mass $\MHpm$ and assuming $\sTheta = 0.5 $. }
\label{fig:crs}
\end{figure}

There are two main sources of SM background.
The first source is the SM production of di-bosons $ZV$, with the $Z$ boson decaying into two OSSF leptons, and another vector boson, $V = W^\pm$ or $Z$, decaying into two jets, i.e., $p\, e^- \to j\, e^-\, Z\, V \to j\, e^-\, (l^+l^-)\, (j j)$. We label it as B1.
The second source is  Z+jets production, with the $Z$ boson decaying leptonically, and two jets produced from QCD radiation, i.e., $p\, e^- \to j\, e^-\, Z\, jj \to j\, e^-\, (l^+l^-)\, j j$. We label it as B2.
Compared with B2, B1 is a pure QED production, and its production cross section is consequently much smaller. However, since the two jets have the invariant mass of a W boson, it is an irreducible background. This is not the case for B2 which can therefore be suppressed more efficiently.

Other background processes with two OSSF leptons which do not have the invariant mass of the $Z$ boson may also contribute to the background but they can be effectively rejected and are hence neglected here.

For the event selection, we firstly apply the following pre-selection:
\begin{enumerate}
\item  At least 3 jets with $p_T > 20$ GeV;
\item At least 3 leptons with $p_T > 10$ GeV; the charges and flavors of the leptons are required to be ($e^-, \mu^+, \mu^-$) or ($e^-, e^+, e^-$);
\item No b-jet with $p_T > 20$ GeV;
\end{enumerate}

In order to construct meaningful kinematical observables,  the two leptons from the Z decay and the two jets from the W decay must be identified.
If the OSSF pair of leptons is $\mu^+\mu^-$, there is no ambiguity. If there are 2 electrons in the final state, the OSSF pair $e^+ e^-$ with an invariant mass closest to the $Z$ boson mass will be identified as $\lZ1$ and $\lZ2$ and the third will be $\eFwd$. Among the first three leading jets, the di-jet pair with invariant mass closest to the W boson boson is considered to be $\jW1$ and $\jW2$, while the remaining jet is regarded as $\jFwd$.

After the pre-selection cuts and identification of $\eFwd,\, \lZ1,\, \lZ2$ and $\jFwd,\, \jW1,\, \jW2$, the following 34 kinematical observables are reconstructed and input into the TMVA package~\cite{TMVA2007} to perform the Boosted Decision Trees (BDT) analysis.
\begin{enumerate}[label*=\arabic*.]

\item global observables:
\begin{enumerate}[label*=\arabic*.]
\item the missing energy $\met$;
\item the scalar sum of the transverse momentum $p_T$ of all jets $H_T$.
\end{enumerate}

\item observables for the forward objects:
\begin{enumerate}[label*=\arabic*.]
\item $p_T$ and the pseudorapidity $\eta$ of $\eFwd$ and $\jFwd$: $p_T(\eFwd)$, $\eta(\eFwd)$, $p_T(\jFwd)$, $\eta(\jFwd)$;
\item $p_T$, $\eta$ and invariant mass $M$ of the system of $\eFwd$ and $\jFwd$: $p_T(\eFwd+\jFwd)$, $\eta(\eFwd+\jFwd)$, and $M(\eFwd+\jFwd)$;
\item the pseudorapidity difference $\Delta\eta$ and the azimuthal angle difference $\Delta\phi$ between $\eFwd$ and $\jFwd$: $\Delta\eta(\eFwd, \jFwd)$, $\Delta\phi(\eFwd, \jFwd)$.
\end{enumerate}

\item observables for the final state $Z$ system:
\begin{enumerate}[label*=\arabic*.]
\item $p_T$ and $\eta$ of $\lZ1$ and $\lZ2$: $p_T(\lZ1)$, $\eta(\lZ1)$, $p_T(\lZ2)$, $\eta(\lZ2)$;
\item $p_T$, $\eta$ and $M$ of the system of $\lZ1$ and $\lZ2$: $p_T(\lZ1+\lZ2)$, $\eta(\lZ1+\lZ2)$, and $M(\lZ1+\lZ2)$;
\item $\Delta\eta$ and $\Delta\phi$ between $\lZ1$ and $\lZ2$: $\Delta\eta(\lZ1, \lZ2)$, $\Delta\phi(\lZ1, \lZ2)$.
\end{enumerate}

\item observables for the final state $W$ system:
\begin{enumerate}[label*=\arabic*.]
\item $p_T$ and $\eta$ of $\jW1$ and $\jW2$: $p_T(\jW1)$, $\eta(\jW1)$, $p_T(\jW2)$, $\eta(\jW2)$;
\item $p_T$, $\eta$ and $M$ of the system of $\jW1$ and $\jW2$: $p_T(\jW1+\jW2)$, $\eta(\jW1+\jW2)$, and $M(\jW1+\jW2)$;
\item $\Delta\eta$ and $\Delta\phi$ between $\jW1$ and $\jW2$: $\Delta\eta(\jW1, \jW2)$, $\Delta\phi(\jW1, \jW2)$.
\end{enumerate}

\item observables for the final state $Z+W$ system:
\begin{enumerate}[label*=\arabic*.]
\item $p_T$, $\eta$ and $M$ of the system of the reconstructed W and Z bosons: $p_T(Z+W)$, $\eta(Z+W)$, and $M(Z+W)$;
\item $\Delta\eta$ and $\Delta\phi$ between the reconstructed W and Z bosons: $\Delta\eta(Z, W)$, $\Delta\phi(Z, W)$.
\end{enumerate}

\end{enumerate}

Fig.~\ref{fig:inputObs} shows the kinematical distributions of some input observables at the FCC-eh with an unpolarized electron beam for the signal benchmark point with $\MHpm$ = 600 GeV (red) and for the SM backgrounds B1 (green) and B2 (blue) after applying the pre-selection cuts.

\begin{figure}[h]
\includegraphics[scale=0.065]{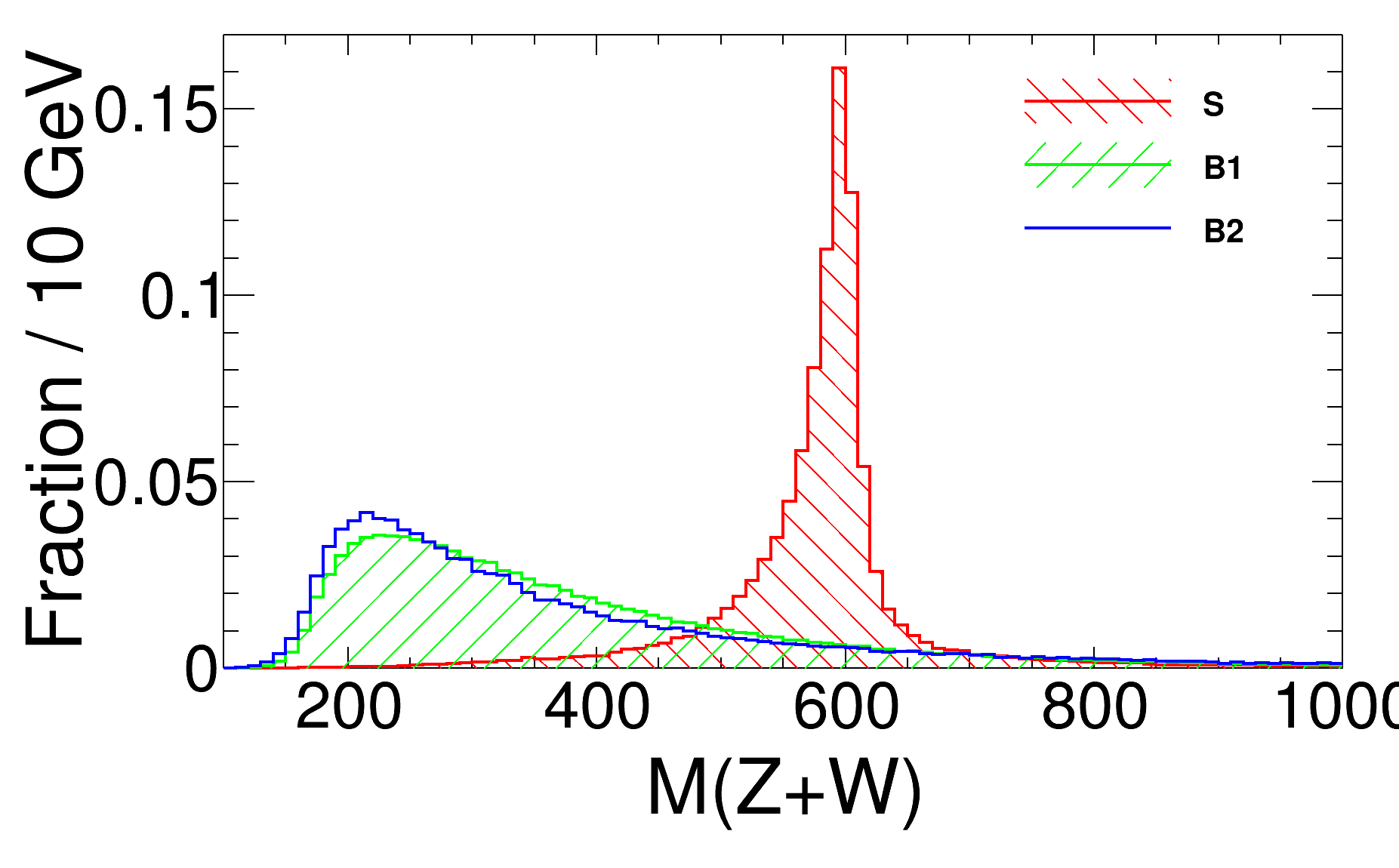}
\includegraphics[scale=0.065]{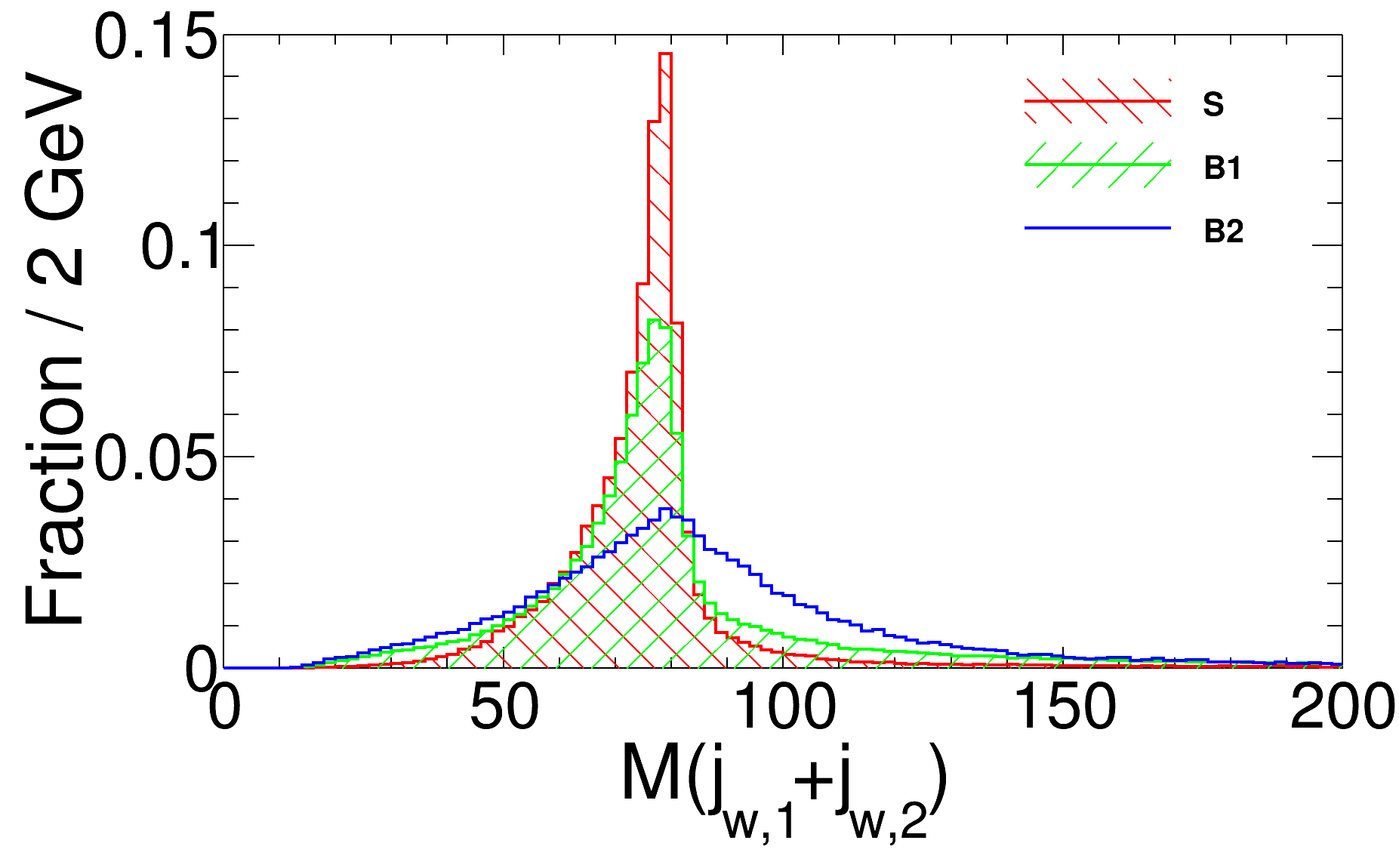}
\includegraphics[scale=0.065]{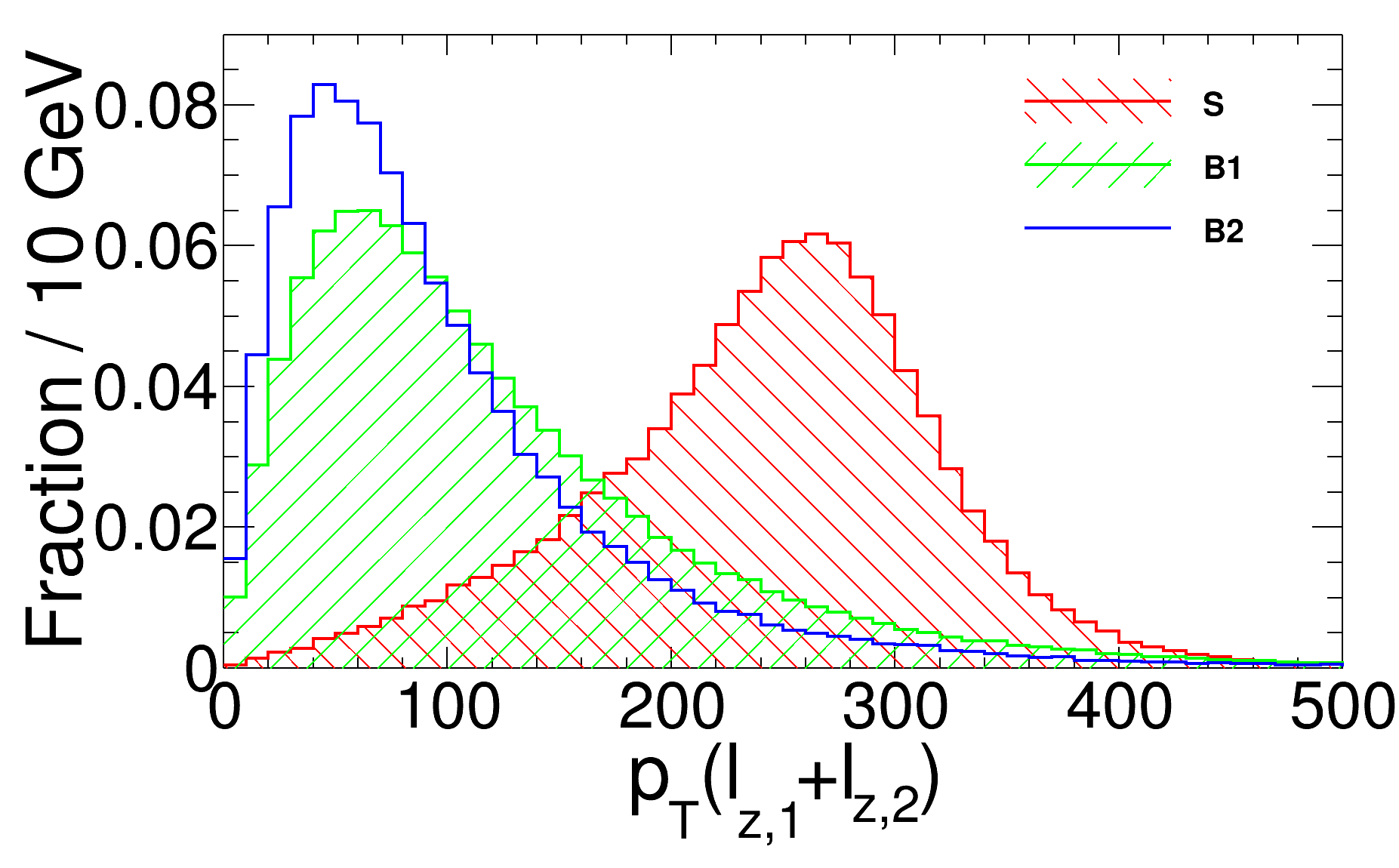}
\includegraphics[scale=0.065]{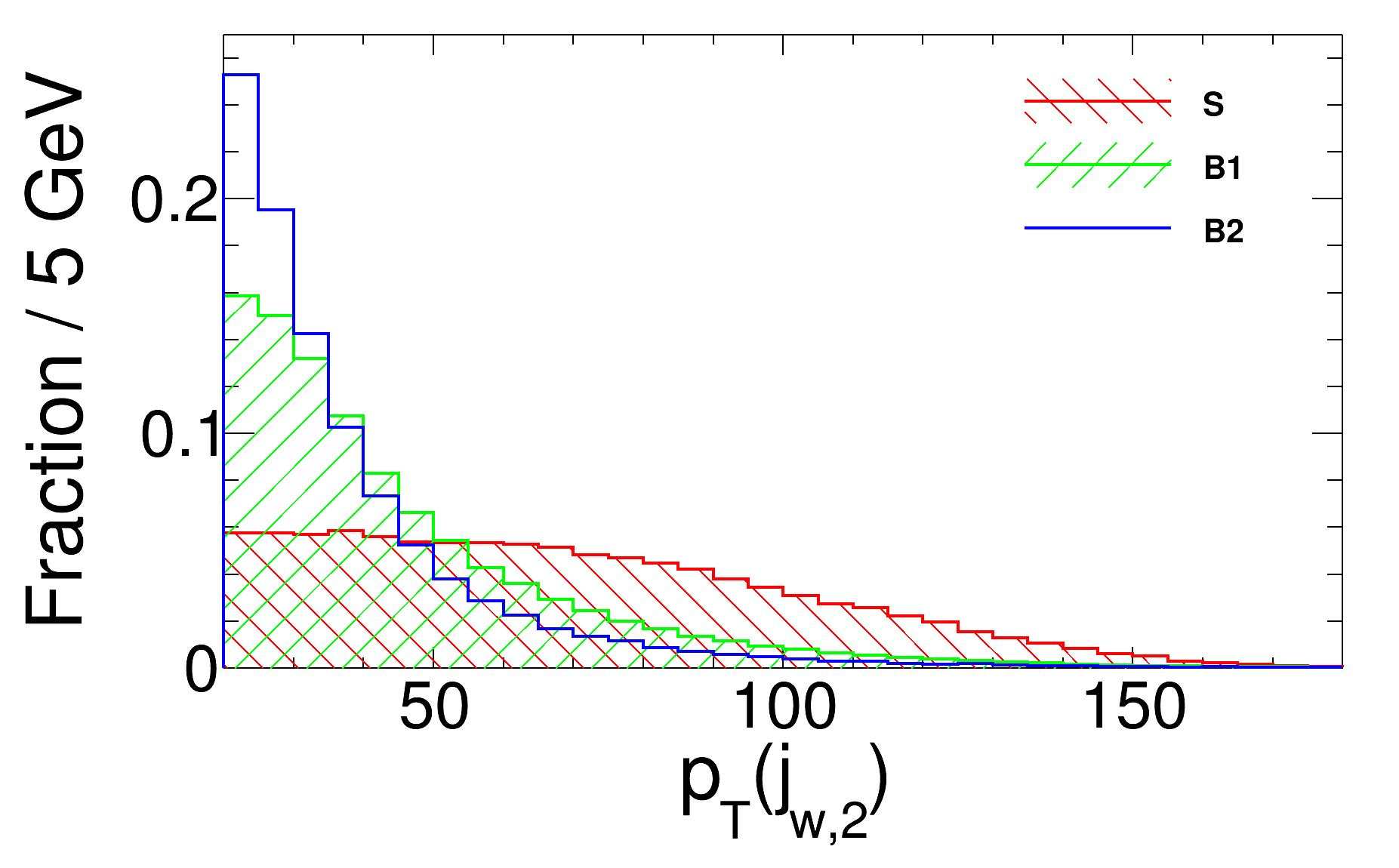}
\includegraphics[scale=0.065]{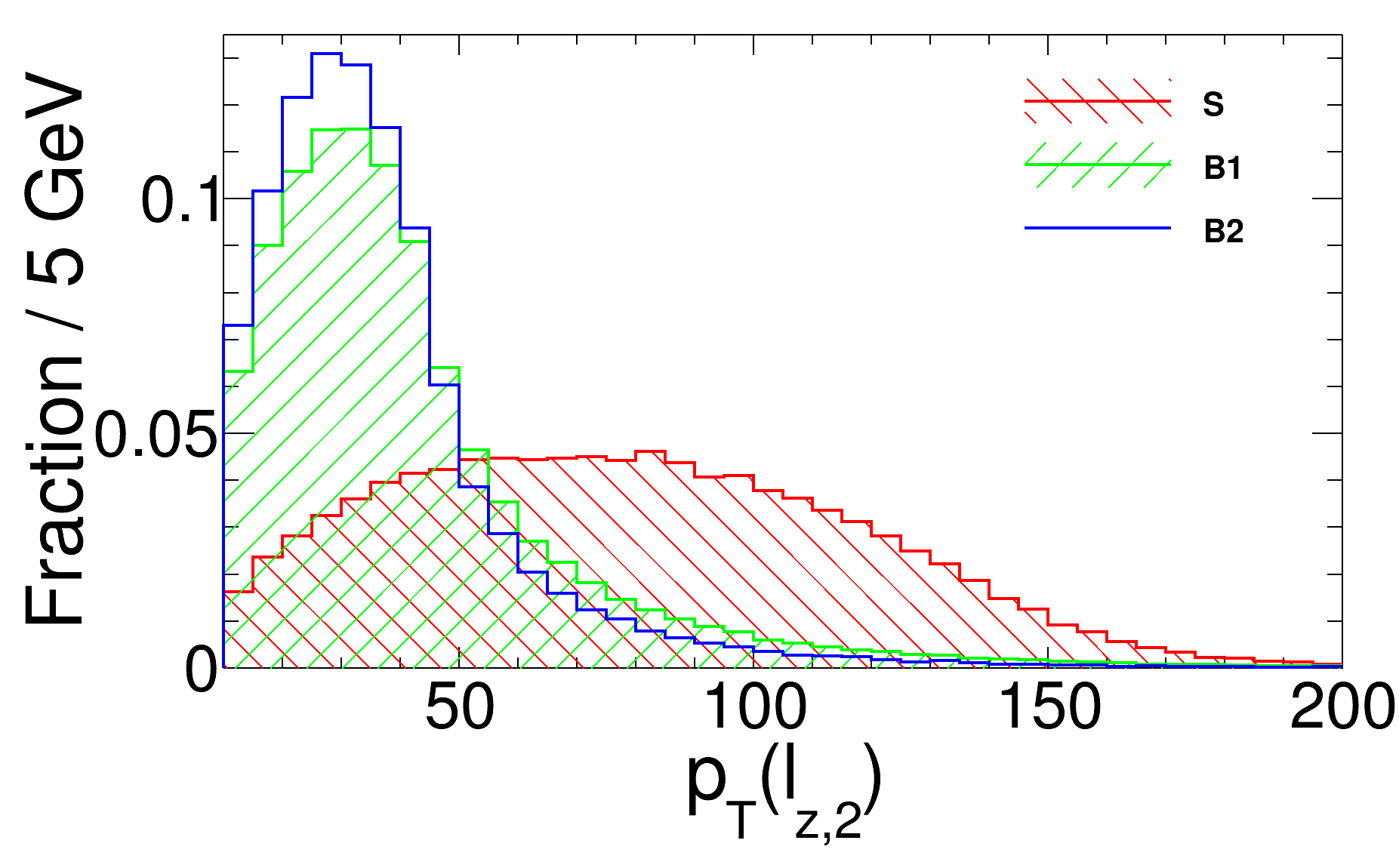}
\includegraphics[scale=0.065]{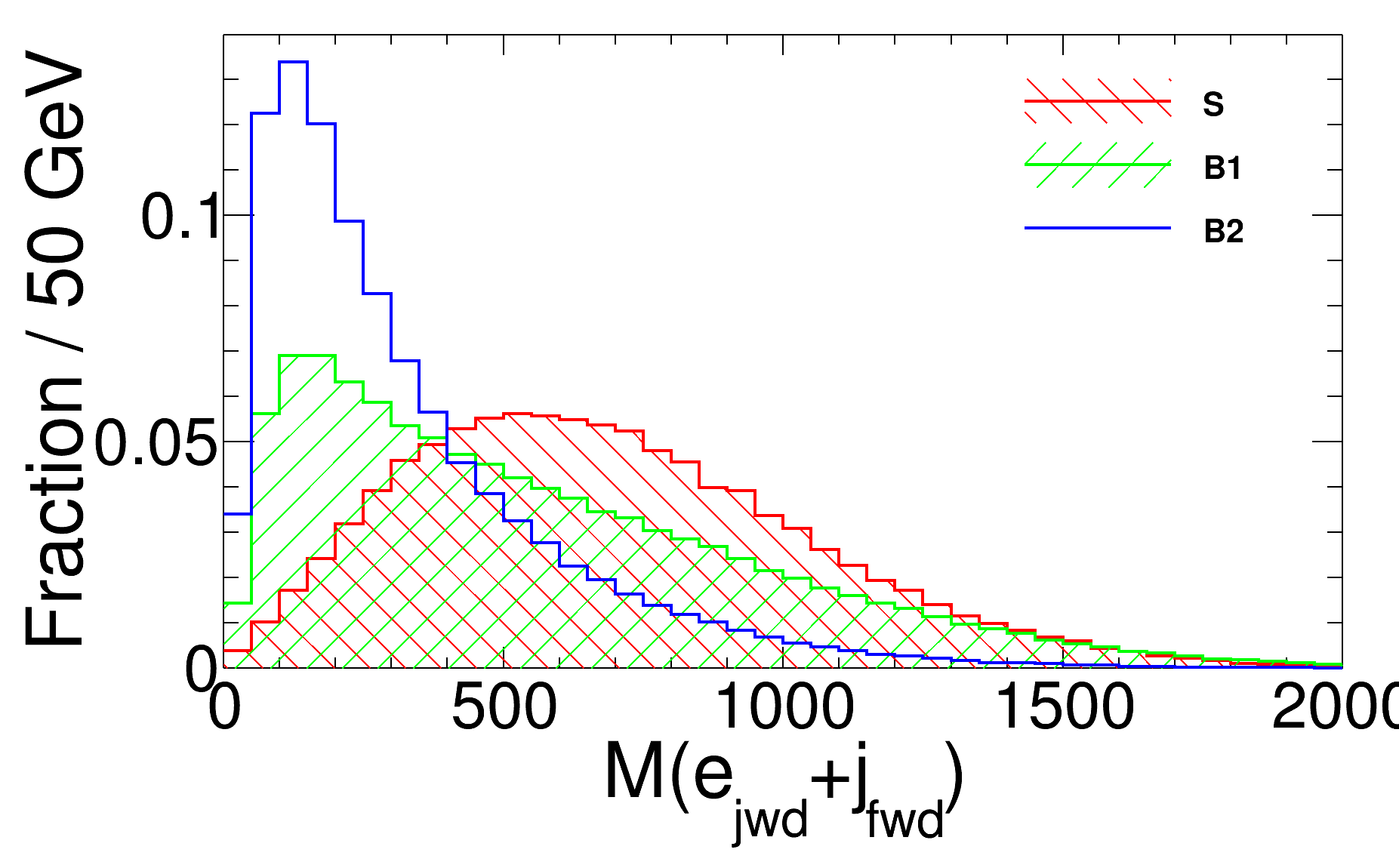}
\includegraphics[scale=0.065]{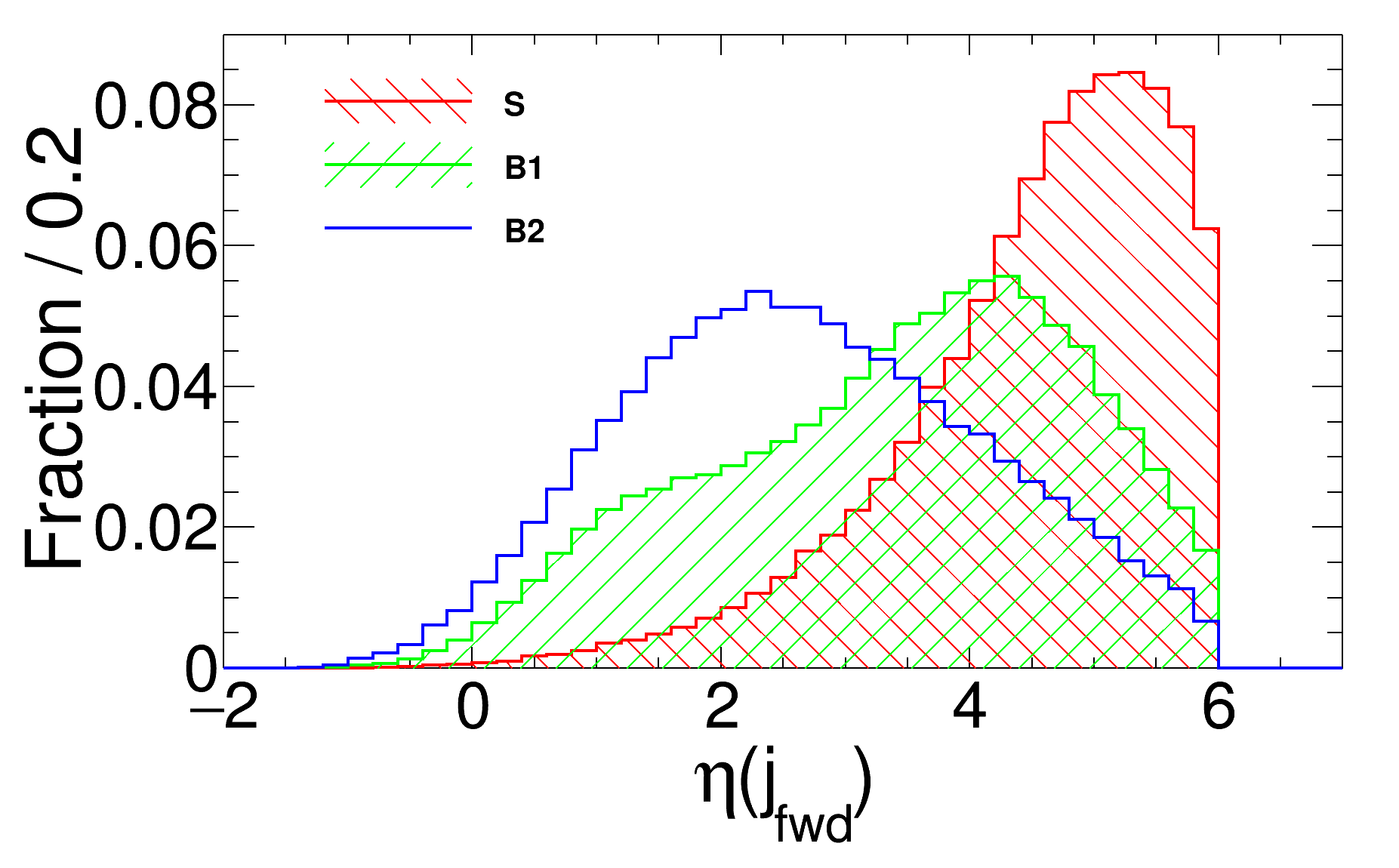}
\includegraphics[scale=0.065]{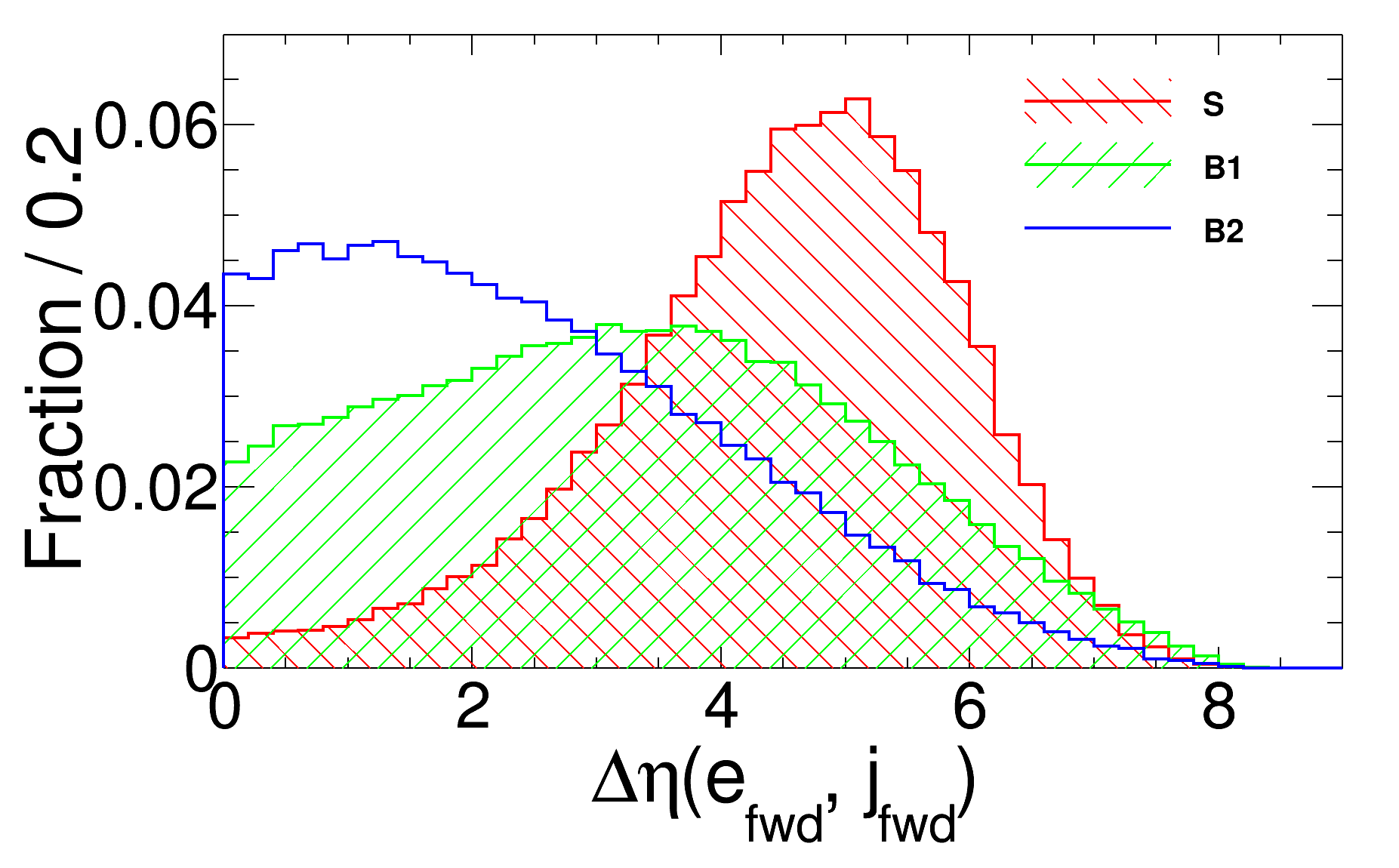}
\includegraphics[scale=0.065]{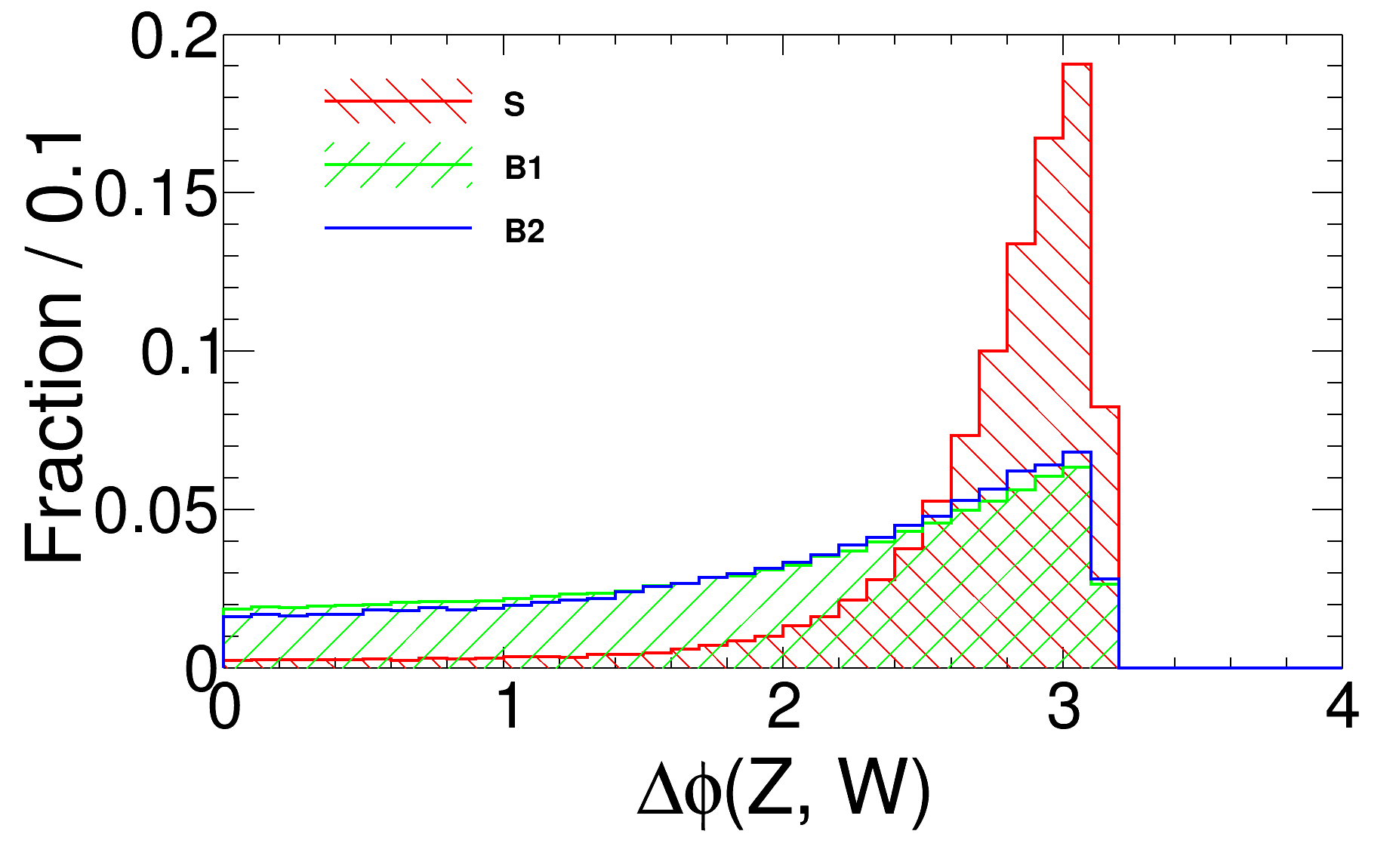}
\includegraphics[scale=0.065]{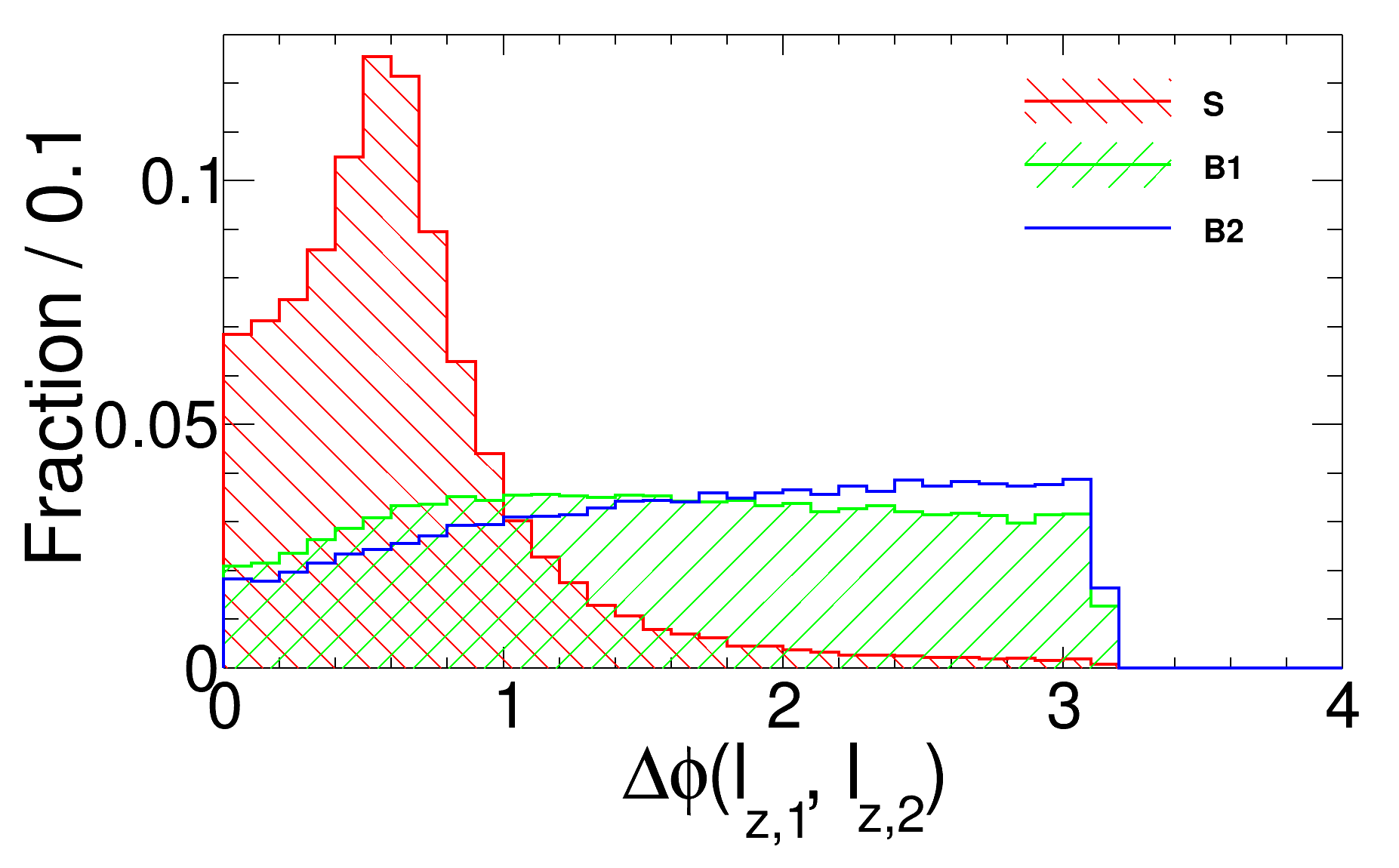}
\caption{ Kinematical distributions of some input observables for signal S (red) with benchmark $\MHpm$ = 600 GeV, and for the SM background B1 (green) and B2 (blue) after applying the pre-selection cuts at the FCC-eh with the unpolarized electron beam.}
\label{fig:inputObs}
\end{figure}

\section{Results}
\label{sec:reuslts}

The observables are input to the TMVA package to perform the BDT analysis.
For the benchmark point $\MHpm$ = 600 GeV, the training and test process reveals that the most useful observables in order of their ranking are $M(Z+W)$, $\eta(\jFwd)$, $p_T(\lZ1+\lZ2)$, $\eta(\eFwd)$, $\Delta\phi(\lZ1, \lZ2)$, $\Delta\eta(\eFwd, \jFwd)$, $M(\jW1+\jW2)$, $\eta(\lZ2)$, $\eta(\jW1)$, $p_T(\jW2)$, $\eta(\lZ1)$, $p_T(\jFwd)$.

In Fig.~\ref{fig:BDT}, we show the distribution of the BDT response for the signal benchmark point with $\MHpm$ = 600 GeV (red), and for the SM background B1 (green) and B2 (blue).
A cut on the  BDT response is chosen to maximize the signal significance.
Assuming an integrated luminosity of $1~\mathrm{ab}^{-1}$ at the FCC-eh (LHeC) with unpolarized electron beam, the cut-flow Table~\ref{tab:sigVSbg} shows the number of events remaining at different stages of the analysis for the signal with $\MHpm$ = 600 (200) GeV and $\sTheta = 0.5$, and for the backgrounds B1 and B2.

\begin{figure}[h]
\includegraphics[scale=0.13]{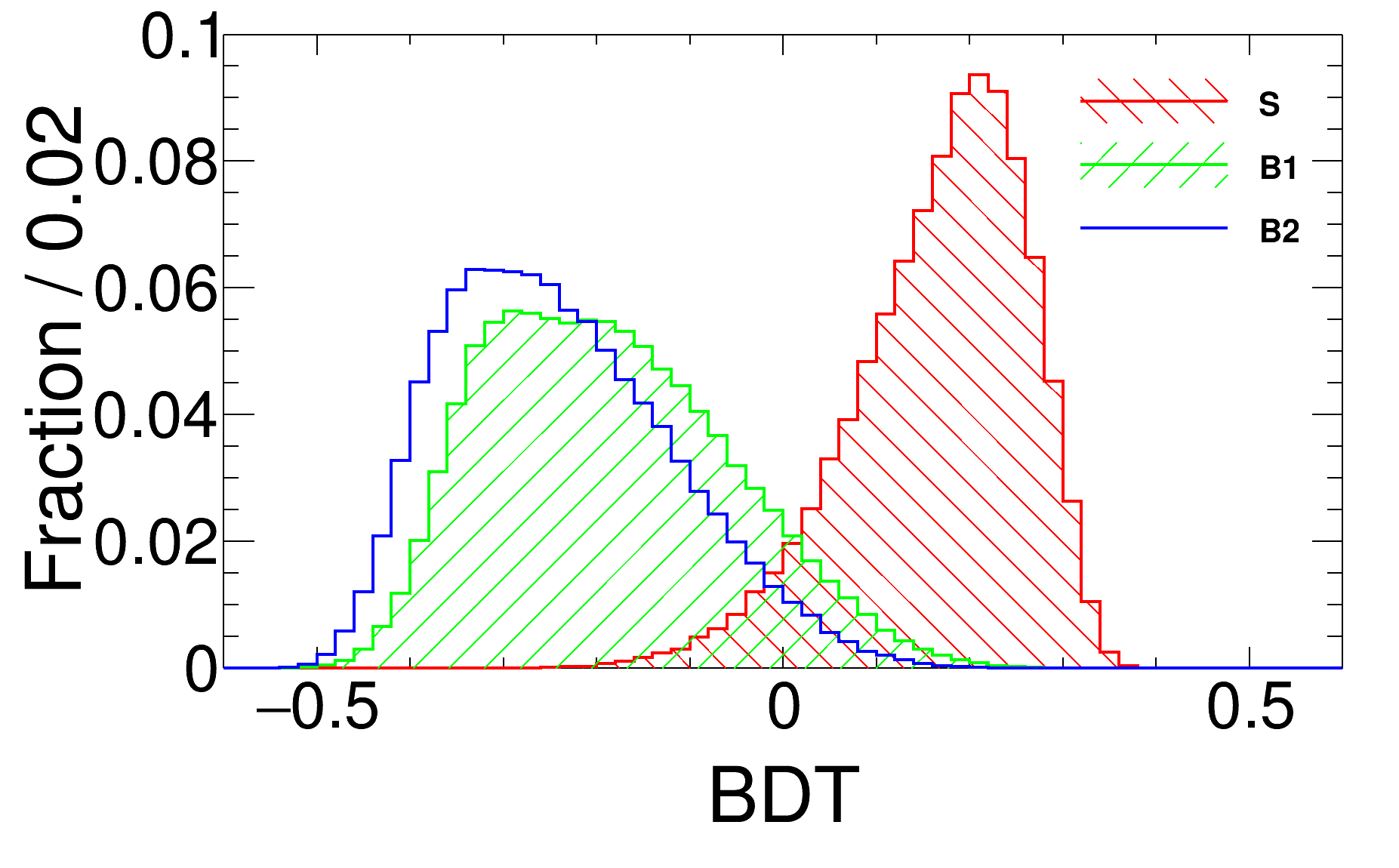}
\caption{ Distribution of BDT response for the signal S (red) with benchmark $\MHpm$ = 600 GeV, and for the SM background B1 (green) and B2 (blue) after applying the pre-selection cuts at the FCC-eh with the unpolarized electron beam.}
\label{fig:BDT}
\end{figure}

The signal significance, 
$\sigma_{stat}$, of the potential signal is evaluated as
\begin{equation}
\sigma_{stat} =
\sqrt{2 [(N_s+N_b) {\rm ln}(1+\frac{N_s}{N_b}) - N_s ] }
\label{eqn:sgf1}
\end{equation}
where $N_s$ ($N_b$) are the expected number of events for signal (background).

Taking into account a systematic uncertainty of $\sigma_b$ in the evaluation of the number of background events, Eq.~(\ref{eqn:sgf2}) will be used to evaluate the significance:

\begin{eqnarray}
\sigma_{stat+syst} =
\Bigg[ 2 \bigg( &(&N_s + N_b) {\rm ln} \frac{(N_s+N_b)(N_b+\sigma_b^2)}{N_b^2+(N_s+N_b)\sigma_b^2} \nonumber \\
 &-& \frac{N_b^2}{\sigma_b^2} {\rm ln}(1+ \frac{\sigma_b^2 N_s}{N_b(N_b+\sigma_b^2) } )\, \bigg)\, \Bigg]^{1/2}
\label{eqn:sgf2}
\end{eqnarray}

For the signal benchmark point at the FCC-eh, the statistical significance is found to be 13.5$\sigma$.
It is difficult to estimate the systematic uncertainties on the background. However, since the signal is well reconstructed as a narrow resonance over a smooth background, a data driven method using the sideband distributions can be used to constrain the level of background. Assuming a systematic uncertainty of $\sigma_b = 10\% \times N_b$ on the background, the signal significance reduces to 12.7$\sigma$.

\begin{table}[h]
\begin{tabular}{c| ccc| ccc}
\hline
\hline
     & \multicolumn{3}{c|}{FCC-eh} & \multicolumn{3}{c}{LHeC} \\
Cuts & S & B1 & B2 & S & B1 & B2 \\
\hline
initial           & 260   & $1.09 \times 10^4$ & $1.52\times 10^5$  & 220 & 531 & $3.19\times 10^4$ \\
Pre-selection     & 102   & 751 & 6442 &  13 &  11 & 148 \\
${\rm BDT}$ &  47.7 & 1.7 & 1.7  &   4.9 &   0.2 &   0.8 \\
\hline
$\sigma_{stat}$ & \multicolumn{3}{c|}{13.5} & \multicolumn{3}{c}{3.35}  \\
$\sigma_{stat+10\% syst}$ & \multicolumn{3}{c|}{12.7} & \multicolumn{3}{c}{3.32}  \\
\hline
\hline
\end{tabular}
\caption{Cut-flow table at the FCC-eh (LHeC) for the signal benchmark point with $\MHpm$ = 600 (200) GeV and $\sTheta = 0.5$, and the SM background B1 and B2. The numbers of events correspond to an integrated luminosity of $1~\mathrm{ab}^{-1}$ with unpolarized electron beam. The optimized BDT cut is ${\rm BDT}> 0.189 (0.119)$ for the FCC-eh (LHeC). The signal significances with 0\% and 10\% systematic uncertainty on background are presented
in the last two rows.}
\label{tab:sigVSbg}
\end{table}

Since the production cross section $\sigma (p\, e^- \to j e^- \Hpm)$ is proportional to $\sin^2 \theta_H$, these results can be reinterpreted in terms of limits on the parameter $\sTheta$. For the benchmark point of $\MHpm = 600$ GeV at the FCC-eh, the $\sTheta$ values corresponding to significances of 5$\sigma$, 3$\sigma$, and 2$\sigma$, assuming 10\% systematic uncertainty on the background, are found to be 0.26, 0.19 and 0.15, respectively.

\begin{figure}[h]
\includegraphics[scale=0.13]{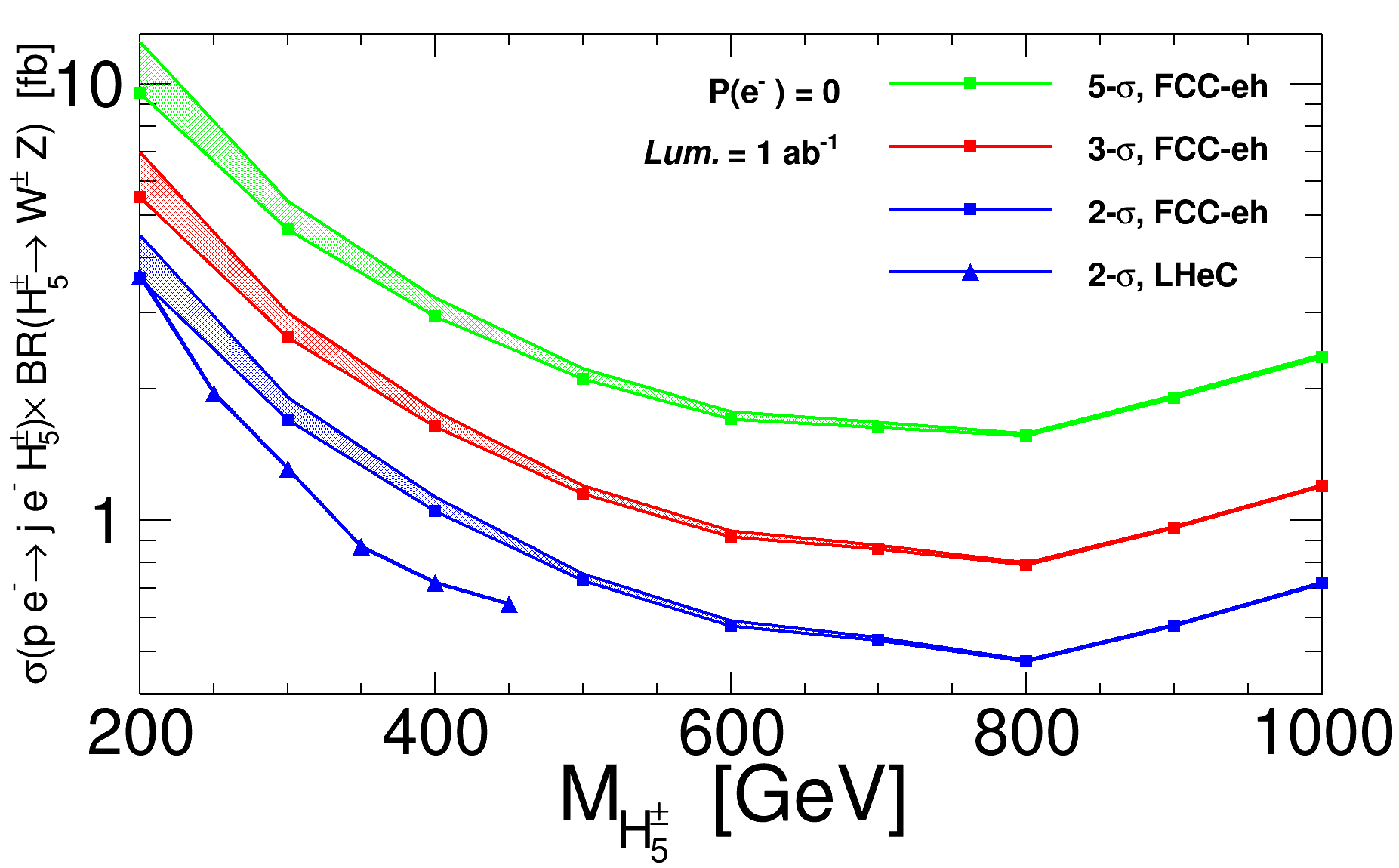}
\caption{The significance contour bands in the plane of production cross section times branching ratio $\sigma (p\, e^- \to j e^- \Hpm) \times {\rm BR}(\Hpm \to Z\, W^\pm) $ vs. $\MHpm$,  for the FCC-eh and LHeC with unpolarized electron beams and luminosity of 1 $\iab$. For each band,
the bottom (top) of the shaded region denotes the significance curve with 0\% (10\%) systematic uncertainty on the background. }
\label{fig:sgf_crs}
\end{figure}

Fig.~\ref{fig:sgf_crs} shows, as a function of $\MHpm$, the cross section times branching ratio $\sigma (p\, e^- \to j e^- \Hpm) \times {\rm BR}(\Hpm \to Z\, W^\pm)$ which can lead to an observation with significance of $2\sigma$, $3\sigma$ and $5\sigma$, assuming  1 $\iab$ of integrated luminosity at FCC-eh. The 2$\sigma$ sensitivity at the LHeC is shown on the same plot, even though the center-of-mass energy of the collider is different.
For the heavy mass points at the FCC-eh and for all the mass points at the LHeC, due to the low number of background events expected after the final BDT cut, a 10\% systematic uncertainty on the background has a negligible effect on the sensitivity of the measurement.
For the benchmark 600 GeV point at the FCC-eh, considering 10\% systematic uncertainty on the background, the cross sections corresponding to the 2, 3, 5-$\sigma$ significances are 0.59, 0.95, 1.78 fb, respectively. At the LHeC with 10\% systematic uncertainty on the background, for the 
200 GeV benchmark point, the cross sections corresponding to the 2-$\sigma$ significance is 3.69 fb.

\begin{figure}[h]
\includegraphics[scale=0.13]{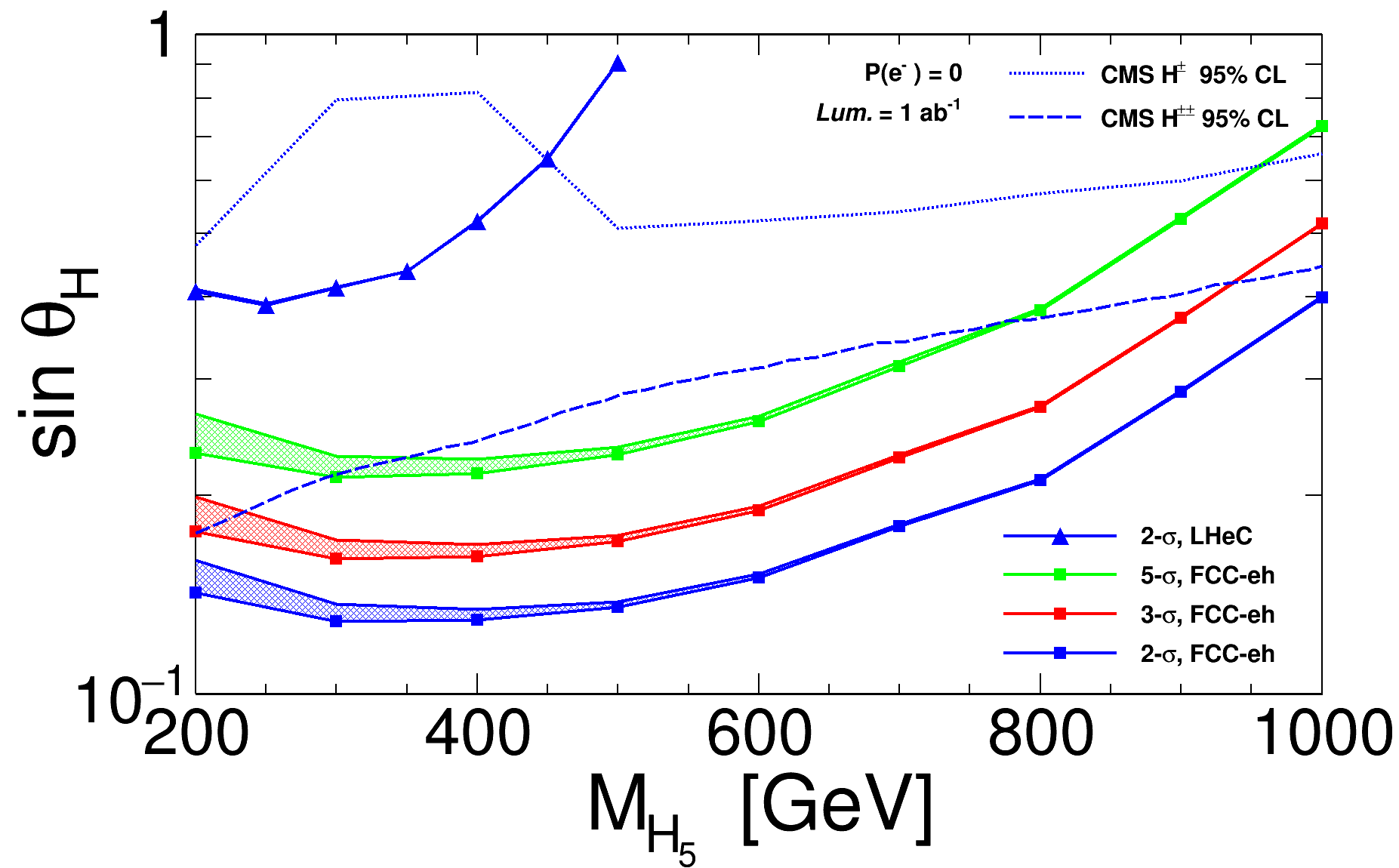}
\caption{ The significance contour bands in the plane of $\sTheta$ vs. $\MH$  for the FCC-eh and LHeC with unpolarized electron beams and luminosity of 1 $\iab$. 
For each band,
the bottom (top) of the shaded region denotes the significance curve with 0\% (10\%) systematic uncertainty on the background. The blue dotted curve gives the 95\% CL limit on the singly charged $\Hpm$ searches at the CMS from Ref.~\cite{GM_Boundary_H5P_CMS}, while the blue dashed curve denotes the 95\% CL limit on the doubly charged $\Hpmpm$ searches at the CMS from Ref.~\cite{GM_Boundary_H5PP_b}.}
\label{fig:sgf_sinTheta}
\end{figure}

Fig.~\ref{fig:sgf_sinTheta} shows the significance contour bands in the plane of $\sTheta$ vs. the five-plet mass $\MH$ for the FCC-eh and LHeC, with unpolarized electron beams and luminosity of 1 $\iab$.
Also shown are the current 95\% CL limits on the singly charged $\Hpm$ searches \cite{GM_Boundary_H5P_CMS} and on the doubly charged $\Hpmpm$ searches~\cite{GM_Boundary_H5PP_b} obtained by the CMS Collaboration.
At the FCC-eh with 10\% systematic uncertainty on the background, the 2 (5)-$\sigma$ limits on the model parameter $\sTheta$ are found to be 0.15 (0.26) for the benchmark 600 GeV mass.
For the benchmark 200 GeV mass point at the LHeC, with 10\% systematic uncertainty on the background the 2-$\sigma$ limits on the $\sTheta$ is 0.41.
Compared with the current CMS limits from the singly charged Higgs searches,
based on 15.2 $\ifb$ of data at 13 TeV, the LHeC 2-$\sigma$ limits are still stronger for the lower masses, while the FCC-eh 2-$\sigma$ limits are much stronger for all masses.
The current doubly charged Higgs searches by CMS,
based on 35.9 $\ifb$ of data at 13 TeV, obtain similar limits for 200 GeV and for 1000 GeV masses to those of the FCC-eh. However, the CMS limits are much weaker for masses around 500 GeV.
It is worth emphasizing that we have assumed degenerate masses for $H_5^{\pm\pm}$ and $H_5^\pm$ here, which may not be the case in a more generic model.

As shown in the Fig.~\ref{fig:crs}, at the FCC-eh for a given mass, a -80\% (+80\%) polarization of electron beam increases (decreases) the production cross section of the signal by a factor of about 10\% compared with the case of an unpolarized beam.
It is found that with the same beam polarizations the cross section of backgrounds B1 and B2 will also increase (decrease) by factors of about 10\% and 25\%, respectively.
Moreover, we find that the kinematical distributions of some input observables such as $\eta(\eFwd)$, $p_T(\eFwd)$, $\Delta\eta(\eFwd, \jFwd)$, $\Delta\phi(\eFwd, \jFwd)$, $p_T(\eFwd+\jFwd)$,  are quite different in the two cases. It is therefore not possible to simply scale the cross sections to infer the limits with polarized beams.
For the benchmark $\MHpm$ = 600 GeV, after performing the full analysis with simulation of both the signal and background data in the polarized electron beam cases, we find at the FCC-eh with 1 $\iab$ luminosity, the 2-$\sigma$ limits on the  $\sTheta$ change only from 0.152 in the case of unpolarized beam to 0.157 (0.148) in the cases of -80\% (+80\%) polarization.
Therefore, beam polarization has a very limited effect on the sensitivity of signal for this study.

\section{Conclusions}
\label{sec:Summary}

We develop the search strategy for the singly charged 5-plet Higgs in the Georgi-Machacek model at the ep colliders.
The charged Higgs are produced by vector boson fusion process,
$p\, e^- \to j\, e^-\, \Hpm$,
and followed by the decays of $\Hpm \to Z\, W^{\pm} \to (l^+ l^-)\, (jj)$.
With a detector simulation, we adopt the BDT method to perform the multivariate analysis and extract the potential signal from the background.
Assuming 10\% uncertainty on the background, at the FCC-eh with an unpolarized electron beam and an integrated luminosity of 1 $\iab$,
we find the 2, 3, and 5-$\sigma$ limits on the production cross section times branching ratio $\sigma (p\, e^- \to j e^- \Hpm) \times {\rm BR}(\Hpm \to Z\, W^\pm)$ and on the model parameter $\sTheta$ for charged Higgs masses in the range 200 to 1000 GeV.
The 2-$\sigma$ limits at the LHeC are also presented. 
The effects of electron beam polarization are investigated and found to be small for this study.
Compared with the present limits obtained from the singly charged Higgs searches at the LHC,
the LHeC center-of-mass energy does not allow a competitive test of the GM model, 
while the FCC-eh limits are much stronger for all masses.
The FCC-eh has therefore a very good potential to search for the charged Higgs with coupling to the vector bosons and for testing the GM model.

\begin{acknowledgments}
We thank Max Klein, Uta Klein, Peter Kostka, Satoshi Kawaguchi, Masahiro Kuze, and Monica D'Onofrio for the data simulation and helpful communications. We appriciate the comments from other members in the LHeC / FCC-eh Higgs and BSM physics study groups.
K.W. also wants to thank Christophe Grojean for his help.
H.S. is supported by the National Natural Science Foundation of China (Grant No. 11675033);
K.W. by the International Postdoctoral Exchange Fellowship Program (No.90 Document of OCPC, 2015);
G.A. by NSERC, Canada.
\end{acknowledgments}





\end{document}

%